# Competing Constraints on Superconductivity in Thick FeSe films

Ya-Xun He,[1] Xing-Jian Liu,[1] Qun Wang,[1] Ting Chen,[1] Hassan Ali,[1] Jia-Ying Zhang,[1] Bao-Juan Kang,[1] Zheng Zhang, [2†] Jun-Yi Ge[1,3,4*]

[1] Materials Genome Institute, Shanghai University, 200444 Shanghai, China

[2] College of Ocean Science and Engineering, Shanghai Maritime University, 201306 Shanghai, China

[3] Department of Physics and Shanghai Key Laboratory for High Temperature Superconductors, Shanghai University, 200444 Shanghai, China

[4] Institute for Quantum Science and Technology, Shanghai University, 200444 Shanghai, China

*Email: junyi_ge@t.shu.edu.cn

†Email: zhangzheng9027@gmail.com

## Abstract

**Superconducting films emerge from the complex interplay of multiple growth parameters, making their optimization challenging. In iron-based superconductors, compressive strain is known to enhance the transition temperature ($T_c$) of FeSe films, yet reported $T_c$ values vary widely even on identical substrates, indicating factors beyond strain are critical. Here, we develop a high-throughput off-center pulsed laser deposition strategy that transforms plume inhomogeneity into combinatorial FeSe film libraries with continuous gradients in lattice parameter, composition, and disorder. We discover that the maximum $T_c$ does not coincide with the plume center but can shift off-center, revealing a competition between favorable $c$-axis expansion, stoichiometry, and defect scattering. Systematic characterization of 80 thick films (>50 nm), combined with interpretable machine learning, shows that besides the strong correlate of $c$-axis lattice parameter to $T_c$, the stoichiometry and disorder scattering impose critical constraints on the achievable transition temperature, defining a narrow optimization window rather than a simple monotonic relationship. This framework yields $T_c^{onset}$=17.1 K in thick FeSe films and establishes a general framework combining combinatorial synthesis with machine learning to uncover constrained optimization landscapes in complex functional materials.**

## Introduction

FeSe, with the simplest crystal structure and composition among iron-based superconductors, provides an ideal platform for investigating the factors that control superconductivity in this family. Its superconducting transition temperature ($T_c$) is remarkably tunable[1,2], increasing from ~ 9 K in bulk crystals[3] to above 65 K in monolayer films[4], and is highly responsive to external stimuli such as ionic-liquid gating[5-10], high pressure[11-15], and chemical intercalation[16-20]. This extraordinary tunability has made FeSe a model system for studying how lattice, charge, and disorder influence high-$T_c$ superconductivity.

In bulk FeSe, superconductivity is highly sensitive to both stoichiometry and disorder. Even a small excess of Fe can strongly suppress, or eliminate, superconductivity[21], while variations in growth conditions and defect scattering can substantially modify both the superconducting and structural transitions.[2] In films, the situation is more complex because superconductivity is influenced not only by stoichiometry or disorder, but also by substrate-film coupling (interfacial interaction)[22-25], strain, thickness[26,27], and microstructure[28,29]. Monolayer FeSe/SrTiO₃, for example, exhibits dramatically enhanced superconductivity because of strong interface effects that are absent in thicker films. Even in thick FeSe films (>50 nm), however, $T_c$ can exceed the bulk value under ambient pressure, with previous reports reaching ~ 15 K.[1,24,30,31]

For thick FeSe films, many studies have linked enhanced $T_c$ to substrate-induced in-plane compressive strain[32-34], which is often manifested experimentally as an expanded out-of-plane $c$-axis lattice parameter.[1,31,34] CaF₂ is a representative example: FeSe films grown on CaF₂ often exhibit enlarged $c$-axis values and improved superconducting properties. Yet, an important question remains unresolved. Why FeSe films grown on same substrate show widely different $T_c$ values, even in some cases despite comparable thicknesses and similarly expanded $c$-axis values.[1,24,27,31,35,36] Addressing this question requires more than comparing independently optimized samples. Instead, it calls for an experimental platform that can span the relevant parameter space continuously within a unified framework, so that the coupled evolution of strain, composition, and disorder can be tracked simultaneously.

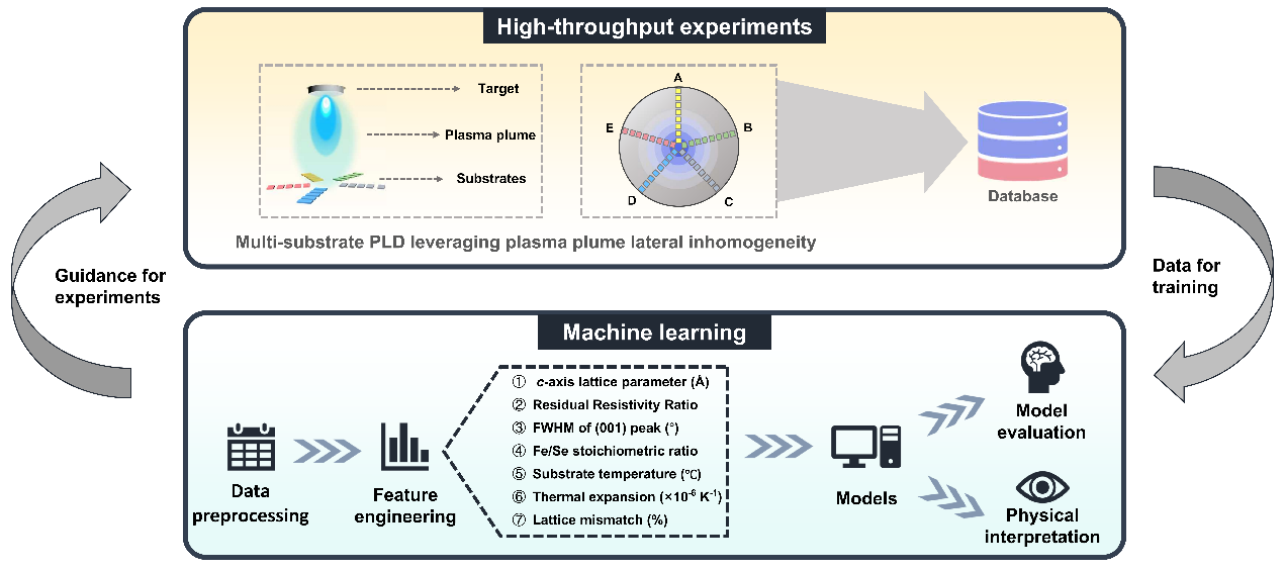

**Figure 1 Workflow of high-throughput experiments and machine learning.** Multi-substrate pulsed laser deposition leveraging plasma plume lateral inhomogeneity generates comprehensive experimental datasets. The machine learning pipeline processes seven key structural and processing parameters through feature engineering, model training, and evaluation to extract the physical insights.

Pulsed laser deposition (PLD) is particularly well suited to this purpose because the intrinsic lateral inhomogeneity of the laser plume naturally generates gradients in deposition flux, kinetic energy, and composition across the substrate.[37,38] Rather than treating this plume inhomogeneity as a limitation, here we deliberately exploit it through an off-center deposition strategy to construct a high-throughput FeSe film library. Combined with interpretable machine learning, this platform allows us to quantify how multiple coupled descriptors govern superconducting performance, as illustrated in Figure 1.

Using this approach, we show that $c$-axis expansion is a robust indicator of a strain-favorable state for superconductivity enhancement in thick FeSe films. However, a large $c$-axis lattice parameter by itself is not sufficient to produce high $T_c$. Instead, the realized superconducting state reflects the coupled effects of strain, stoichiometry, and disorder scattering. In particular, anomalous off-center superconducting maxima reveal that when the plume-center film is initially Fe-rich, the improvement in stoichiometry away from the center can initially outweigh the reduction in $c$-axis expansion. Guided by this coupled optimization landscape, we achieve $T_c^{onset} = 17.1$ K in a 150 nm FeSe/CaF$_2$ film. These results define a finite strain-stoichiometry window for optimizing superconductivity in thick FeSe films and demonstrate a broader strategy for disentangling multidimensional synthesis-structure-property relationships in complex film materials.

## Results

### High-throughput FeSe film library

Figure 2a shows a schematic of the plasma plume during deposition. As illustrated in Figure 2b, because the plume center has higher particle density, higher kinetic energy, and a larger deposition rate than off-center positions, this geometry introduces continuous gradients in growth conditions within a single deposition run. To construct a combinatorial FeSe film library, substrates were placed at different radial positions relative to the PLD plume center (Figure 2c). We used five substrate types：CaF₂, MgO, SrTiO₃, LaAlO₃, and Si—to expand the accessible strain background while simultaneously imposing lateral gradients in composition and defect formation. This strategy produced multiple FeSe films in one run.

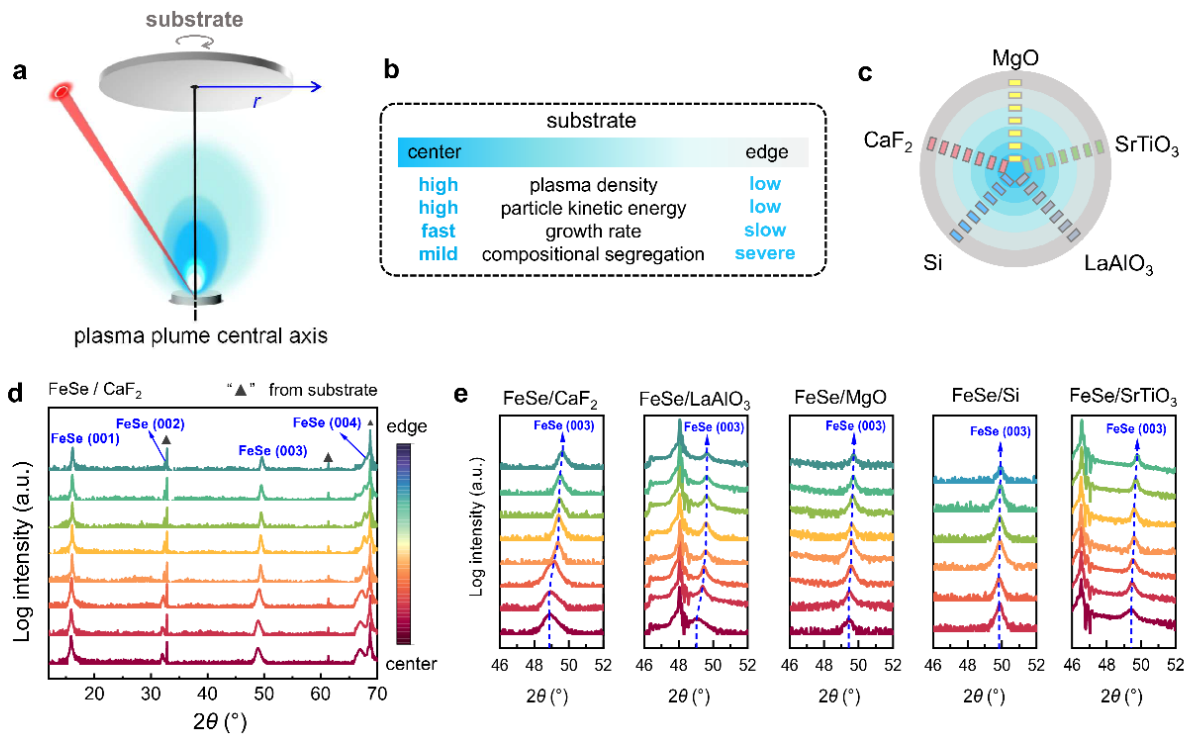

**Figure 2 High-throughput fabrication and characterization of FeSe films. (a)** Schematic of the PLD configuration showing the plasma plume and the rotating substrate holder, with r indicating the distance from the plume center. **(b)** Gradient deposition conditions across the substrate holder, highlighting the systematic change in plasma density, particle kinetic energy, and growth rate, and the compositional segregation from the center to the edge. **(c)** Schematic illustration showing the placement of five different kinds of substrates (colored squares) on the holder. All substrates are aligned along the radial direction. **(d)** Out-of-plane XRD $2\theta$ scans of FeSe/CaF$_2$ films at different radial positions from center to edge. The observed 00l diffraction peaks suggest epitaxy growth of FeSe films. **(e)** Enlarged FeSe (003) peaks for films on different kinds of substrates and radial positions. The systematic change of peak position suggests the successful high-throughput fabrication of FeSe films with different lattice parameters. The dashed lines are guide to eyes.

Figure 2d shows the XRD patterns for representative FeSe films on CaF₂, which confirm the superconducting $\beta$-tetragonal phase structure. X-ray diffraction shows that all films are $c$-axis-oriented and adopt the superconducting tetragonal FeSe phase. Complete XRD patterns for films on other substrates are presented in Supplementary Fig. S1. The (003) peak systematically shifts toward higher angle with increasing radial distance, indicating a gradual decrease in the $c$-axis lattice parameter from the plume center toward the edge (Figure 2e). Among all substrates, FeSe/CaF₂ exhibits the largest $c$-axis lattice parameter, whereas FeSe/Si shows the smallest, immediately suggesting a strong substrate dependence of the strain state (Figure 3a). Here, the $c$-axis lattice parameters are extracted via Nelson-Riley extrapolation[39] (see Supplementary Fig. S2). Energy-dispersive spectroscopy (EDS) further reveals pronounced substrate-dependent stoichiometry evolution. Here, EDS analysis, averaging over 12 data points per sample (see Supplementary Fig. S3), characterized film stoichiometry through [Fe]/[Se] atomic ratios. As shown in Figure 3b, FeSe/Si is strongly Fe-rich and becomes more Fe-rich with increasing radial distance, while FeSe/LaAlO$_3$ remains globally Se-rich. By contrast, FeSe/MgO and FeSe/SrTiO$_3$ stay close to nominal stoichiometry, and FeSe/CaF$_2$ evolves from slightly Fe-rich near the center toward a more stoichiometric state off-center. These contrasting trends indicate that the final [Fe]/[Se] ratio is governed not only by the plume itself, but also by substrate-specific interfacial chemistry and post-deposition diffusion. Film thickness also varies systematically with radial position and is coupled to the lattice state. Taking FeSe/CaF$_2$ as an example (Figure 3c), thicker films near the plume center exhibit larger $c$-axis values, whereas thinner off-center films approach the bulk lattice constant. Although a simple elastic relaxation picture would predict the opposite trend, the present result indicates that under off-center deposition the film thickness, lattice state, and disorder co-evolve with the strong radial gradients in plume flux and particle energetics, and therefore should not be treated as independent variables. Likewise, the full width at half maximum (FWHM) of the (001) peak shows that crystallinity is often optimized not exactly at the plume center, but within an intermediate off-center window (Figure 3d). The residual resistivity ratio (RRR) varies systematically as well, providing a transport-based descriptor of the defect-scattering environment.

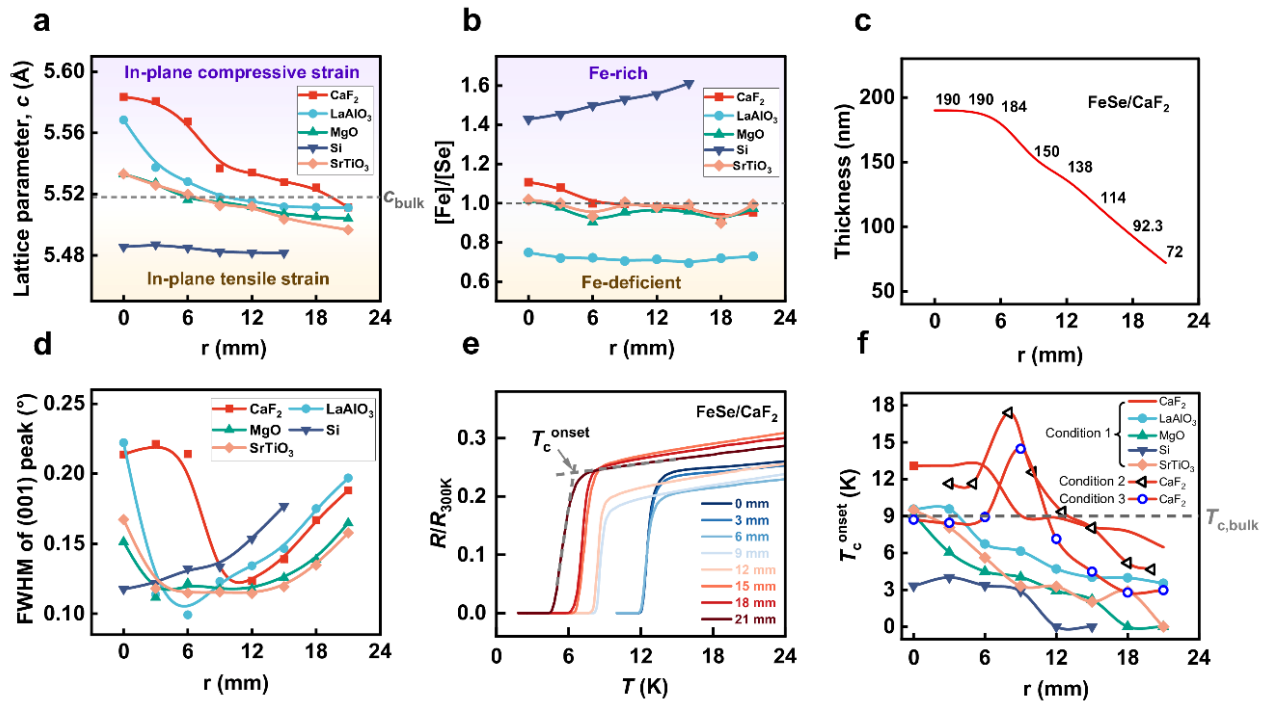

**Figure 3 Characterization of high-throughput FeSe films. (a)** Out-of-plane lattice parameter $c$ as a function of radial position $r$ for FeSe films on CaF₂, LaAlO₃, MgO, Si and SrTiO₃ substrates. The horizontal dashed line marks the $c_{bulk}$ value for bulk FeSe. **(b)** Atomic ratio of [Fe]/[Se] versus $r$ for the same set of substrates. The dashed line denotes the Fe/Se boundary between Fe-rich and Fe-deficient compositions. **(c)** Film thickness as a function of $r$ for FeSe grown on CaF₂. **(d)** FWHM of the FeSe (001) XRD peak as a function of r for different substrates. **(e)** Temperature dependence of the normalized resistivity $R/R_{300K}$ for FeSe/CaF₂ measured at different $r$. **(f)** Radial dependence of the $T_c^{onset}$ for FeSe films. Solid symbols (Condition 1, $T_s$ = 500 °C) correspond to combinatorial growth on CaF₂, LaAlO₃, MgO, SrTiO₃, and Si substrates, while open and filled red symbols (Condition 2, $T_s$ = 300 °C, and Condition 3, $T_s$ = 350 °C) denote additional CaF₂-only runs grown under modified PLD conditions. In most depositions the maximum $T_c^{onset}$ appears near the plume center, but in condition 2 and 3 it shifts to off-center positions, giving rise to non-monotonic $T_c^{onset}$ profiles. The horizontal dashed line marks the $T_c$ of bulk FeSe.

**Radial evolution of superconductivity and off-center optima**

The most direct manifestation of these coupled gradients appears in the superconducting transition. Figure 3e presents representative $R$-$T$ curves for FeSe/CaF$_2$ films fabricated at different radial positions. Here, $T_c^{onset}$ is defined as the temperature at which $R/R_{300\,K}$ begins to deviate from the normal-state linear regime upon cooling. The corresponding $R$-$T$ curves for FeSe films on LaAlO$_3$, MgO, SrTiO$_3$, and Si are provided in Supplementary Fig. S4. The spatial evolution of $T_c^{onset}$ is summarized in Figure 3f. As shown in Figure 3f, we can see that in the initial multi-substrate experiment at $T_s = 500$ °C, the highest $T_c^{onset}$ on each substrate is found near the plume center. At a given radial distance, the general hierarchy is CaF$_2$ > LaAlO$_3$ > SrTiO$_3$ > MgO > Si, consistent with the fact that CaF$_2$ provides the most favorable strain background among the substrates examined here.

However, further CaF$_2$-based combinatorial runs at lower substrate temperatures reveal a more interesting behavior: the maximum $T_c^{onset}$ no longer always occurs at the plume center, but can shift to an off-center position. In these cases, the radial $T_c$ profile becomes non-monotonic, with $T_c$ first increasing and then decreasing away from the center. The overall record $T_c^{onset}$ ~17.1 K in our 150 nm FeSe/CaF$_2$ films is realized at such an off-center optimum, and the corresponding $R$-$T$ data for this optimal sample are shown in Supplementary Fig. S5.

These off-center maxima are highly informative because they cannot be explained by $c$-axis expansion alone. The $c$-axis lattice parameter decreases monotonically with increasing radial distance, a purely strain-dominated picture would predict a monotonic decrease in $T_c$. Instead, the anomalous non-monotonic profiles appear when the plume-center region is initially Fe-rich. Under these conditions, moving off-center reduces the $c$-axis lattice parameter but simultaneously shifts the Fe/Se ratio toward a more favorable window. The initial improvement in stoichiometry can therefore outweigh the negative effect of reduced $c$-axis expansion. Only at larger radial distance, when the strain weakens further and film quality deteriorates, does $T_c$ decrease again. By contrast, when the plume-center film is already close to the optimal Fe/Se balance, $T_c$ decreases monotonically from center to edge, consistent with the simultaneous reduction of $c$ and degradation of growth conditions.

This comparison provides direct experimental evidence that superconductivity in thick FeSe films is not a simple monotonic function of $c$-axis expansion. Rather, the realized $T_c$ emerges from competition among strain-favorable lattice distortion, stoichiometry optimization, and disorder evolution. This point is central: $c$-axis expansion defines a favorable structural background, but whether that background is converted into high $T_c$ depends on the optimization of both composition and defects.

**Multivariate analysis by interpretable machine learning**

To quantify these coupled effects, we expanded the dataset by focusing additional combinatorial growth on CaF$_2$, which offers both the highest $T_c$ values and the broadest superconducting distribution.

By varying substrate temperature, laser fluence, target-to-substrate distance, and pulse number, we obtained a total of 80 FeSe films, all thicker than 50 nm. This thickness constraint excludes the ultrathin regime, where strong interface-specific mechanisms dominate and air sensitivity introduces additional complications.

Seven descriptors were extracted from structural characterization, transport measurements, growth conditions, and substrate properties. The complete dataset is provided in Supplementary Table S1. The distribution of $T_c^{onset}$ across the dataset is shown in Figure 4a and spans nearly 0-20 K. These strongly co-varying variables make it difficult to reliably establish the relative importance of each factor for superconducting performance. We therefore used machine learning not as a replacement for physical interpretation, but as a quantitative tool for ranking coupled descriptors. As is evident from Figure 4b, among Least Absolute Shrinkage and Selection Operator (LASSO), support vector machine (SVM), random forest (RF), and XGBoost regression, the tree-based models performed best, with XGBoost giving the highest cross-validated $R^2$. After Bayesian hyperparameter optimization, the final model achieved $R^2 = 0.98$ on the training set and $R^2 = 0.91$ on the test set (Figure 4c), indicating that it captures robust structure-property relationships across films grown under different conditions and on different substrates.

The built-in XGBoost feature importance and SHAP analysis yield a consistent physical picture. Among all descriptors, the $c$-axis lattice parameter is the most important single feature, followed by RRR, FWHM, and [Fe]/[Se], while substrate thermal expansion coefficient, substrate temperature, and lattice mismatch contribute less directly (Figure 4d). The SHAP analysis further shows that larger $c$-axis values generally shift the predicted $T_c^{onset}$ upward, in agreement with the global experimental trend (Figure 4e).

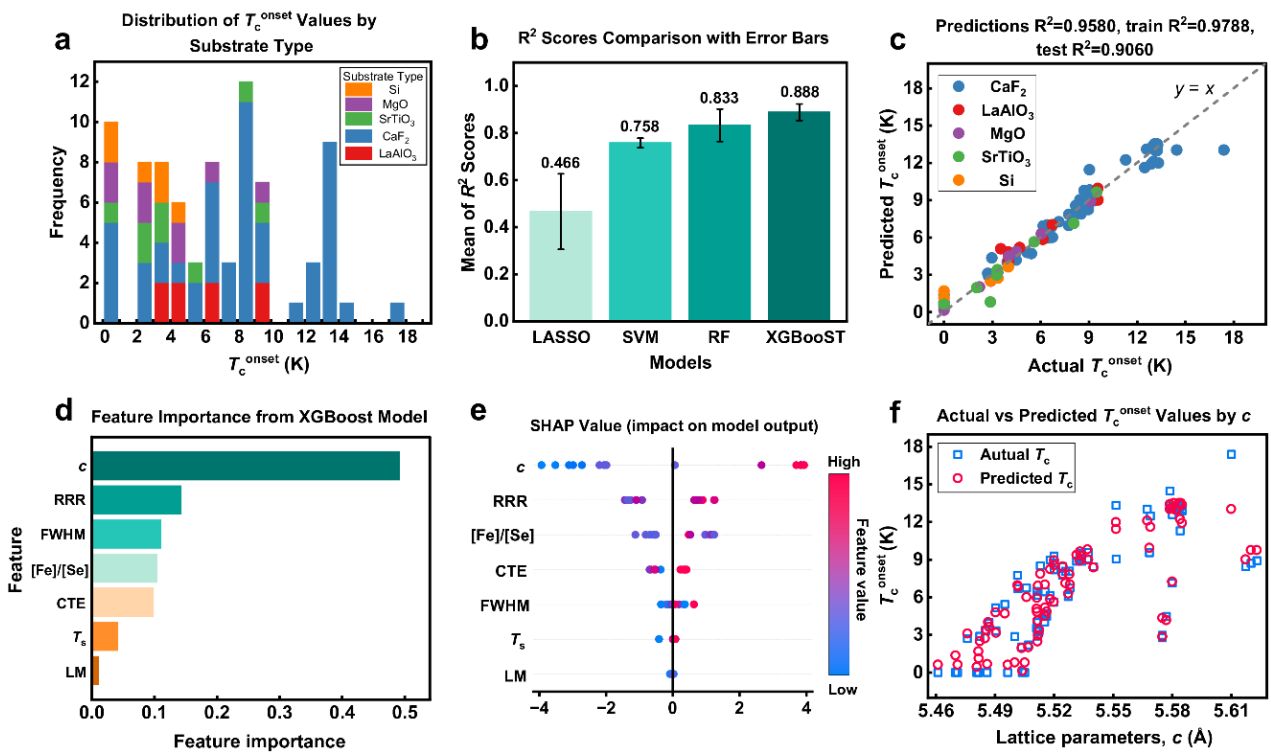

**Figure 4 Machine-learning analysis of the high-throughput FeSe film dataset. (a)** Distribution of the $T_c^{onset}$ across different substrates. **(b)** Comparison of regression-model performance evaluated by mean $R^2$ scores with error bars. **(c)** Predicted versus experimental $T_c^{onset}$ values for the optimized XGBoost model, showing the performance on both training and test sets. **(d)** Feature-importance ranking derived from the XGBoost model based on the F-score. **(e)** SHAP summary plot showing the impact of each descriptor on the model output. **(f)** Experimental and model-predicted $T_c^{onset}$ plotted as a function of the $c$-axis lattice parameter.

Figure 4f further shows that $T_c^{onset}$ increases overall with the $c$-axis lattice parameter across the dataset, although a substantial spread remains among films with similar $c$ values. Higher RRR values also tend to promote higher $T_c$, whereas larger FWHM values are generally unfavorable. The effect of [Fe]/[Se] confirms that composition remains an important constraint on superconducting performance. The non-negligible roles of RRR and [Fe]/[Se] make clear that $c$-axis expansion alone does not fully determine the superconducting state. Rather, $c$ is best understood as the most informative descriptor of a favorable strain background, while the realized $T_c$ still depends on the coupled effects of stoichiometry and disorder scattering.

## Discussion

The global positive correlation between $T_c^{onset}$ and the $c$-axis lattice parameter can be understood in terms of thermoelastic strain. Because the films are grown at elevated temperature, the mismatch in thermal contraction between FeSe and the substrate upon cooling produces residual in-plane strain. When the substrate has a larger thermal expansion coefficient than FeSe, the film experiences in-plane compression during cooling, accompanied by a Poisson-like expansion of the $c$-axis. This provides a simple explanation for the special role of $CaF_2$: compared with conventional oxide substrates, $CaF_2$ combines a relatively favorable lattice match with a larger thermal expansion coefficient, thereby stabilizing a compressive-strain state that promotes $c$-axis expansion and enhanced superconductivity (see also Supplementary Fig. S6). Our highest-$T_c$ sample, with $T_c^{onset} = 17.1$ K and $c = 5.61$ Å, is fully consistent with this picture.

At the same time, the present data clarify why large $c$-axis expansion is not a sufficient condition for high $T_c$. Contrary to the common assumption that maximizing $c$-axis expansion optimizes $T_c$, the anomalous off-center samples provide the clearest evidence: they retain a large $c$-axis parameter, and thus remain within the strain-favorable side of the parameter space, but their superconducting performance is limited when the composition is too Fe-rich (see also Supplementary Fig. S7 for representative radial evolution patterns). These observations directly show that stoichiometry remains a decisive constraint even under favorable strain conditions. In this sense, $c$-axis expansion should be regarded as a key enabling factor, not as a unique control parameter.

The role of disorder is supported most clearly by the machine-learning analysis. RRR carries greater importance than [Fe]/[Se], but this should not be interpreted as meaning that disorder is more fundamental than composition. Rather, RRR is likely an integrated transport descriptor that captures the cumulative consequences of multiple scattering sources, including microstructural inhomogeneity, point defects, and stoichiometry-related disorder. By contrast, the EDS-derived [Fe]/[Se] ratio is a spatially averaged compositional metric and may not fully resolve the local non-stoichiometry most relevant to superconductivity. This explains why RRR can appear more important than [Fe]/[Se] in the model while the direct experimental evidence for stoichiometric suppression remains strong.

From a practical perspective, these results suggest that the design rule for high-$T_c$ thick FeSe films is not simply to maximize $c$-axis expansion, but to access a finite strain-favorable window in which compressive strain, near-stoichiometric composition, and weak disorder scattering are achieved simultaneously. $CaF_2$ has so far proven especially effective because it naturally satisfies these requirements better than most oxide substrates. More generally, substrates or buffer-layer architectures with large thermal expansion coefficients and tolerable lattice mismatch should provide promising routes for further improvement, provided that the beneficial strain can be maintained without introducing excessive disorder or relaxation.

Beyond FeSe, the present work highlights a broader strategy for complex functional materials. Deliberately exploiting spatial growth gradients to generate combinatorial property libraries, and combining these libraries with interpretable machine learning, provides a powerful route to disentangle competing descriptors within multidimensional synthesis spaces. This framework should be broadly applicable to materials in which lattice distortion, composition, and disorder are strongly intertwined in determining electronic, magnetic, or ferroic properties.

## Conclusion

To conclude, we developed a high-throughput off-center PLD platform that transforms intrinsic plume inhomogeneity into a useful tool for mapping coupled growth variables in thick FeSe films. Across a combinatorial library of 80 samples, $c$-axis expansion emerges as the strongest single descriptor of superconducting performance, confirming its role as a robust indicator of a strain-favorable state. At the same time, anomalous non-monotonic $T_c(r)$ profiles reveal that high $T_c$ is realized only when this favorable strain background is combined with near-optimal stoichiometry and sufficiently weak disorder scattering. Guided by this multidimensional understanding, we achieve $T_c^{onset} = 17.1$ K in a 150 nm FeSe/CaF$_2$ film under ambient pressure. More broadly, this work establishes a general route for combining combinatorial thin-film growth with interpretable machine learning to uncover hidden optimization windows and accelerate the design of complex quantum and functional materials.

## Experimental Section

### Materials preparation and characterization

A dense FeSe$_{0.94}$ polycrystalline target ($\rho = 5.35$ g/cm$^3$) for PLD was prepared by high-pressure vacuum sintering. Fe powder ($\geq$ 99.9%) and Se powder ($\geq$ 99.99%) were weighed in an Ar-filled glovebox (O$_2$, H$_2$O < 0.1 ppm), and thoroughly mixed and ground. The mixture was sealed in an evacuated quartz ampoule and subjected to a solid-state reaction: the temperature was slowly ramped to 1080 °C, held for 24 h, and then furnace-cooled to room temperature to obtain a solid product. This product was ground into powder and sintered several times under identical conditions to improve phase homogeneity, yielding a uniform Fe-Se polycrystalline precursor. The precursor powder was then reground, loaded into a graphite die, and hot-pressed under vacuum better than $5 \times 10^{-3}$ Pa with a uniaxial pressure of ~25 MPa at 600 °C for 5 h. After furnace cooling, a highly dense FeSe$_{0.94}$ polycrystalline pellet was obtained and subsequently used as the PLD target. Further details on target preparation have been reported elsewhere[40].

FeSe films were deposited on a series of commercial single-crystal substrates, including CaF$_2$(001) (Hefei CPI Equipment Technology Co., Ltd.), LaAlO$_3$(100), MgO(100), and SrTiO$_3$(100) (all purchased from Hefei Kejing Co., Ltd.), and Si(100) (Suzhou Crystal Silicon Electronic & Technology Co., Ltd.), using a Neocera PLD system equipped with a KrF excimer laser ($\lambda = 248$ nm, COHERENT COMPex 201F). The growth was carried out under a base pressure below $1 \times 10^{-8}$ Torr, with the substrate temperature and laser pulses varying from 300 - 600 °C and 7000 - 10000, respectively, while the target-to-substrate distance, laser energy density, and repetition rate were kept constant at 50 mm, 1.2 ~ 1.8 J·cm$^{-2}$, and 3 Hz.

For characterization, the crystal structure was characterized by $\theta$-$2\theta$ scans using an X-ray diffractometer (XRD; Bruker D2 PHASER). Compositional analysis was performed via energy-dispersive X-ray spectroscopy (EDS; Bruker Quantax 75) coupled to a scanning electron microscope (SEM; Hitachi FlexSEM 1000 II). Film thickness was determined by atomic force microscopy (AFM; Oxford Instruments MFP-3D Origin) step-height measurements after creating substrate-exposing scratches with a quartz needle. Electrical transport properties were measured using the standard four-probe method in a PPMS-DynaCool-14 T system (Quantum Design).

### Machine learning

XGBoost is an open-source library that provides an efficient implementation of the gradient boosting framework[41]. The algorithm adds regression trees during training to minimize prediction errors. Hyperparameters, including n_estimators, max_depth, and learning_rate, are configured before training to optimize the training process, further improving prediction accuracy and reducing overfitting. We applied XGBoost with the Scikit-learn package[42]. Hyperparameters were optimized using Bayesian optimization through the Optuna library[43]. For each hyperparameter set, a five-fold cross-validation resampling technique was employed with stratified sampling to split the train-test sets. The training data was randomly split into five equally sized groups, ensuring each group contained all types of substrates, resulting in five distinct $R^2$ scores for each hyperparameter set. The set with the highest average validation $R^2$ score was selected.

**Acknowledgments**
The study was supported by the National Natural Science Foundation of China (Grant Nos. 12174242). J-Y G also thanks the support by the Program for Professor of Special Appointment (Eastern Scholar) at Shanghai Institutions of Higher Learning.

**Author contributions:**
J.Y.G. conceived the idea and designed the experiments. Y.X.H., T.C., Q.W., K.H.Y., B.J.K. and J.Y.Z. prepared the films and performed the characterization. X.J.L. performed the machine learning with the contributions from Z.Z.. Y.X.H. and J.Y.G. wrote the manuscript. All authors contributed to the discussion and analysis of the data.

**Competing interests:**
The authors declare no competing interests.

**Data and materials availability:**
All data are available in the main text or the supplementary materials.

# Competing Constraints on Superconductivity in Thick FeSe films

Ya-Xun He,[1] Xing-Jian Liu,[1] Qun Wang,[1] Ting Chen,[1] Hassan Ali,[1] Jia-Ying Zhang,[1]
Bao-Juan Kang,[1] Zheng Zhang, [2†] Jun-Yi Ge[1,3,4*]

[1] *Materials Genome Institute, Shanghai University, 200444 Shanghai, China*

[2] *College of Ocean Science and Engineering, Shanghai Maritime University, 201306 Shanghai, China*

[3] *Department of Physics and Shanghai Key Laboratory for High Temperature Superconductors,*
*Shanghai University, 200444 Shanghai, China*

[4] *Institute for Quantum Science and Technology, Shanghai University, 200444 Shanghai, China*

*Email: junyi_ge@t.shu.edu.cn

†Email: zhangzheng9027@gmail.com

**This PDF file includes:**

Figs. S1 to S7
Tables S1

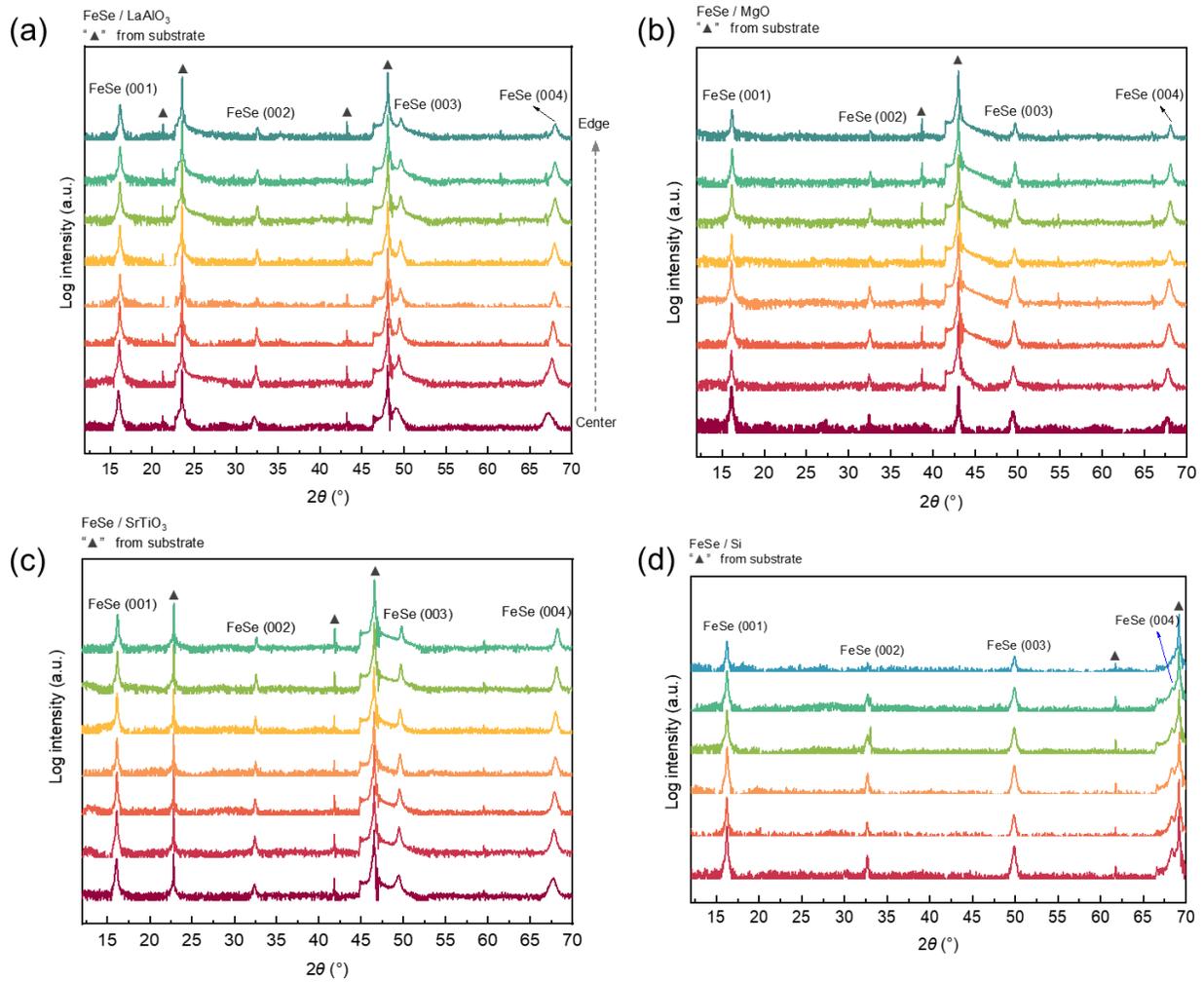

**Fig. S1. Out-of-plane XRD scans confirm highly *c*-axis-oriented FeSe growth on different substrates across plume radial positions.** Out-of-plane XRD 2$\theta$-$\theta$ scans of (a) FeSe/LaAlO$_3$, (b) FeSe/MgO, (c) FeSe/SrTiO$_3$ and (d) FeSe/Si films measured at different radial positions from plume center to edge. Strong FeSe (00l) reflections are observed for all substrates and radial positions, demonstrating highly *c*-axis-oriented growth across the entire plume. These results are consistent with the FeSe/CaF$_2$ data shown in the main text.

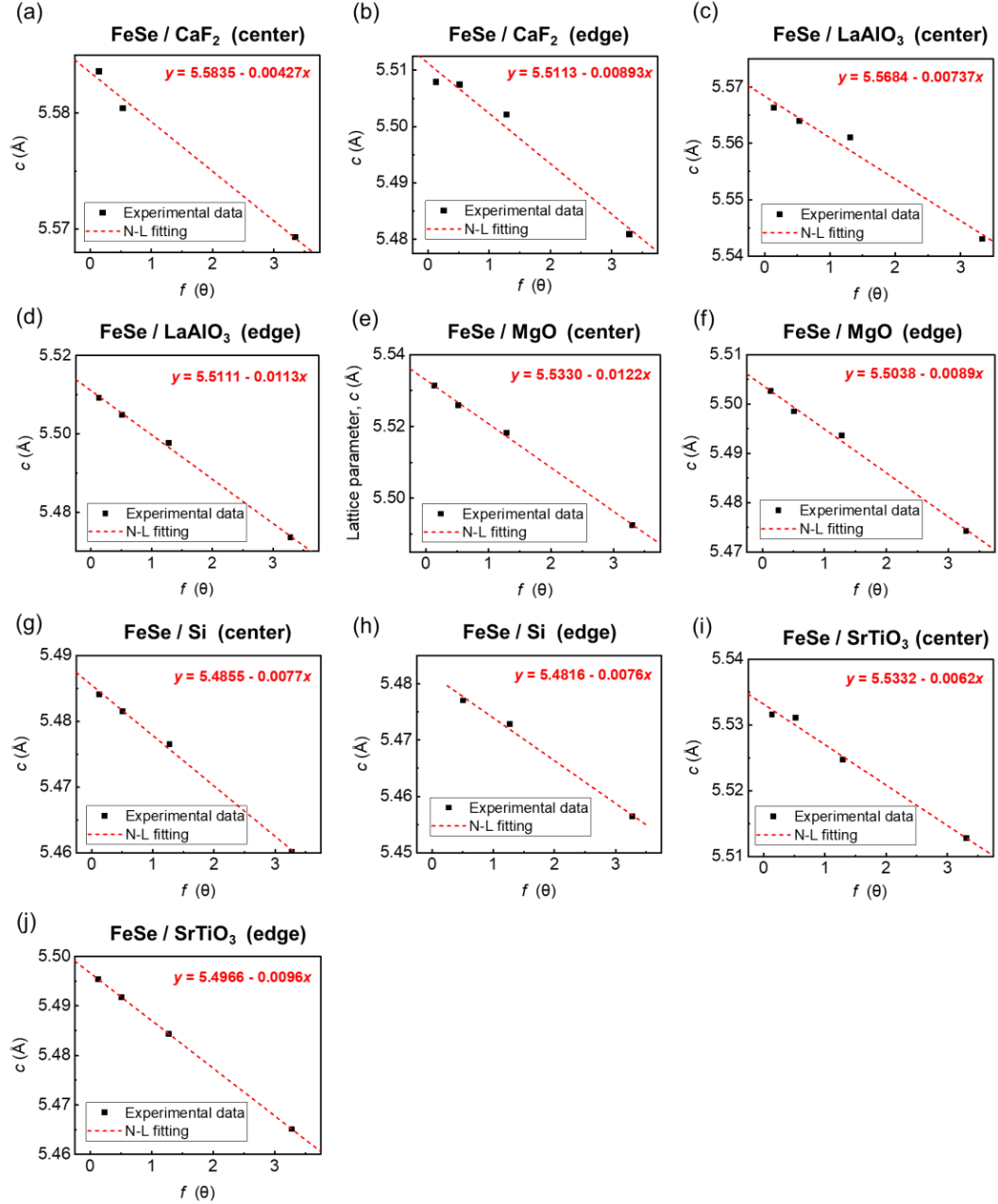

**Fig. S2**. **Nelson-Riley analysis refines the _c_-axis lattice parameter of FeSe films on different substrates and plume positions.** _c_ values extracted from multiple (00*l*) diffraction peaks are plotted as a function of the Nelson-Riley function $f(\theta) = \frac{1}{2}\left(\frac{cos^2\theta}{sin\theta} + \frac{cos^2\theta}{\theta}\right)$ and linearly extrapolated to $f(\theta) \rightarrow$ 0 to obtain the refined _c_-axis lattice parameters for FeSe/CaF$_2$, FeSe/LaAlO$_3$, FeSe/MgO, FeSe/Si, and FeSe/SrTiO$_3$ films at center and edge positions. In this work, the c-axis lattice parameter of every FeSe film is determined using this Nelson-Riley extrapolation of multiple (00l) XRD peaks, which minimizes θ-dependent systematic errors by compensating for peak-position shifts.

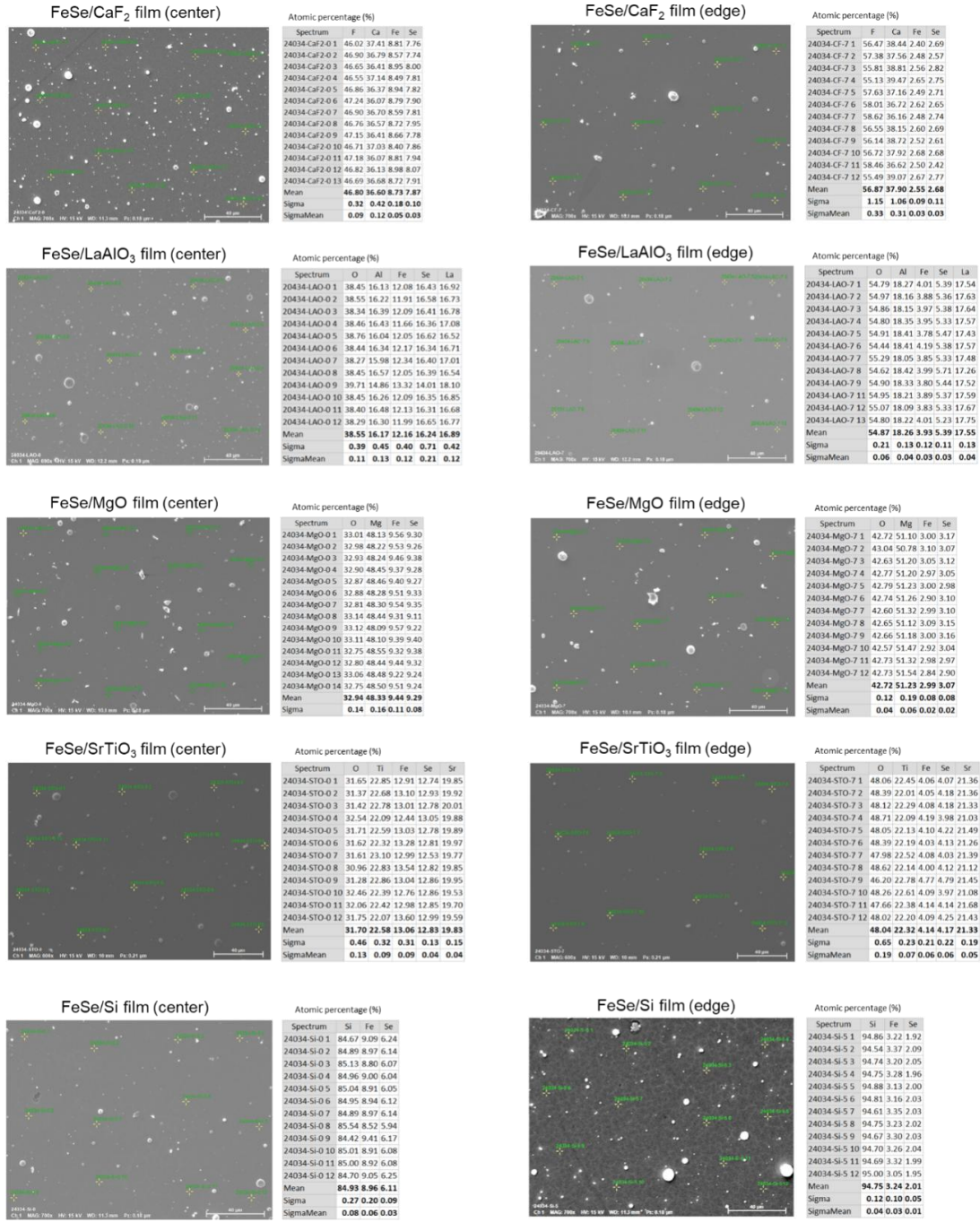

**Fig. S3**. **SEM–EDS analysis reveals the local compositions of FeSe films across different substrates and plume positions.** Representative SEM images and EDS spot analyses for FeSe films grown on CaF₂, LaAlO₃, MgO, SrTiO₃ and Si substrates at center and edge positions. Yellow markers in the SEM micrographs indicate the EDS sampling points used for quantitative analysis, and the accompanying tables summarize the atomic percentages of Fe, Se and substrate elements at each point. Each panel shows a SEM image with EDS sampling points (yellow crosses) overlaid, together with the quantitative atomic composition measured at each site. The data demonstrate compositional uniformity together with substrate-specific elemental signatures in typical regions of FeSe films grown on CaF₂, LaAlO₃, MgO, SrTiO₃ and Si. All film compositions in this work were obtained using the same data-acquisition and processing procedures.

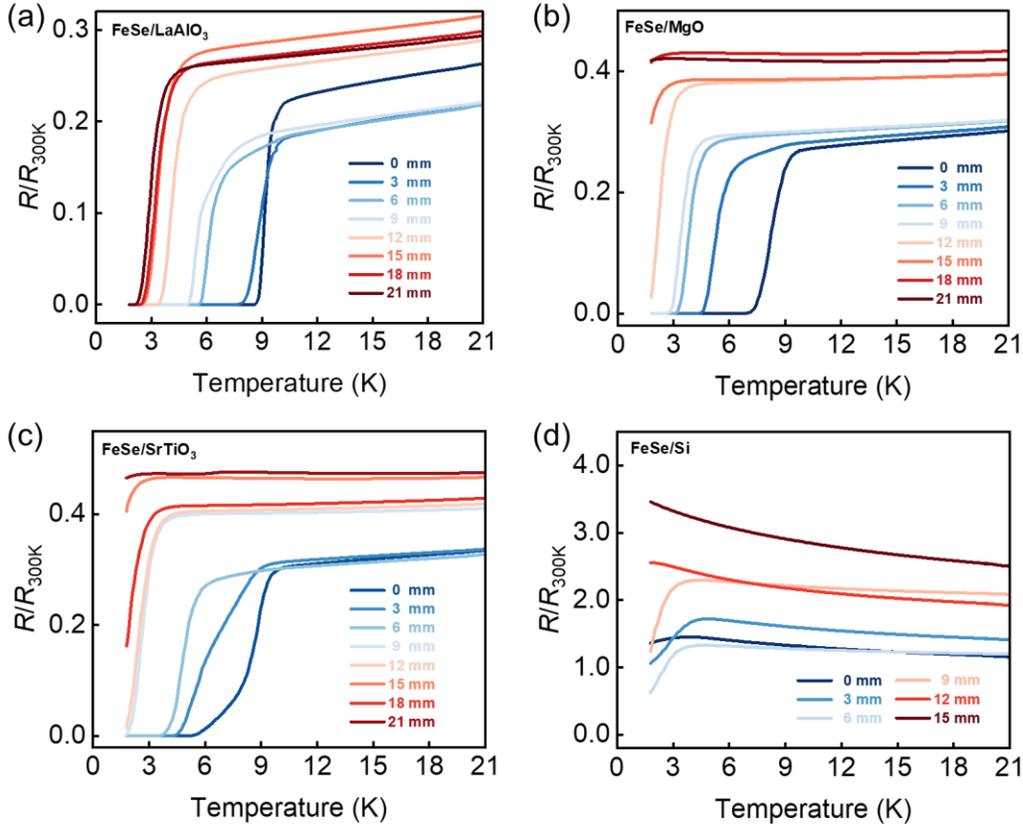

**Fig. S4. Superconducting transitions in FeSe films depend on substrate and off-center deposition position.** Temperature dependence of the normalized resistance $R/R_{300K}$ for FeSe films deposited at various off-center positions on **(a)** LaAlO$_3$, **(b)** MgO, **(c)** SrTiO$_3$ and **(d)** Si substrates. Within each panel, "0 mm" denotes the sample grown at the center of the substrate holder, directly aligned with the plasma plume, whereas larger distances (e.g., 3 mm, 6 mm, …) indicate that the substrate was placed away from the plume center during film deposition. Each panel therefore compiles $R$-$T$ curves from FeSe films deposited at different spatial positions on the same type of substrate. As shown, the $T_c$ exhibits a strong dependence on the off-center distance $r$ for FeSe films grown on different substrates, reflecting the pronounced spatial variation of growth conditions within the PLD plume.

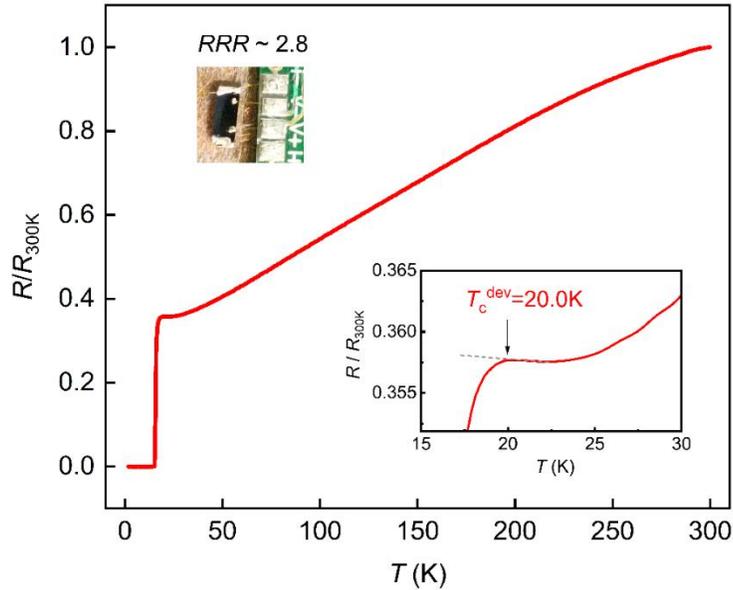

**Fig. S5. Normalized resistance of the best-performing FeSe film reveals the highest $T_c$ in the library.** Temperature dependence of the normalized resistance $R/R_{300K}$ for the FeSe film with the highest superconducting transition temperature in our sample library, measured from 1.8 to 300 K. The inset shows a magnified view of the superconducting transition region, where $T_c^{dev}$=20.0 K marks the temperature at which the resistance first starts to deviate from the high-temperature linear behavior. The residual resistivity ratio (RRR ≈ 2.8) is indicated in the main panel, and a photograph of the patterned electrode configuration used for transport measurements is also provided. This high-$T_c$ FeSe film exhibits a clear superconducting transition: the resistance begins to deviate from the linear normal-state behavior at $T_c^{dev}$ = 20.0 K, whereas the superconducting onset temperature $T_c^{onset}$ quoted in the main text is determined more strictly from the intersection between the extrapolated normal-state line and the steepest part of the transition.

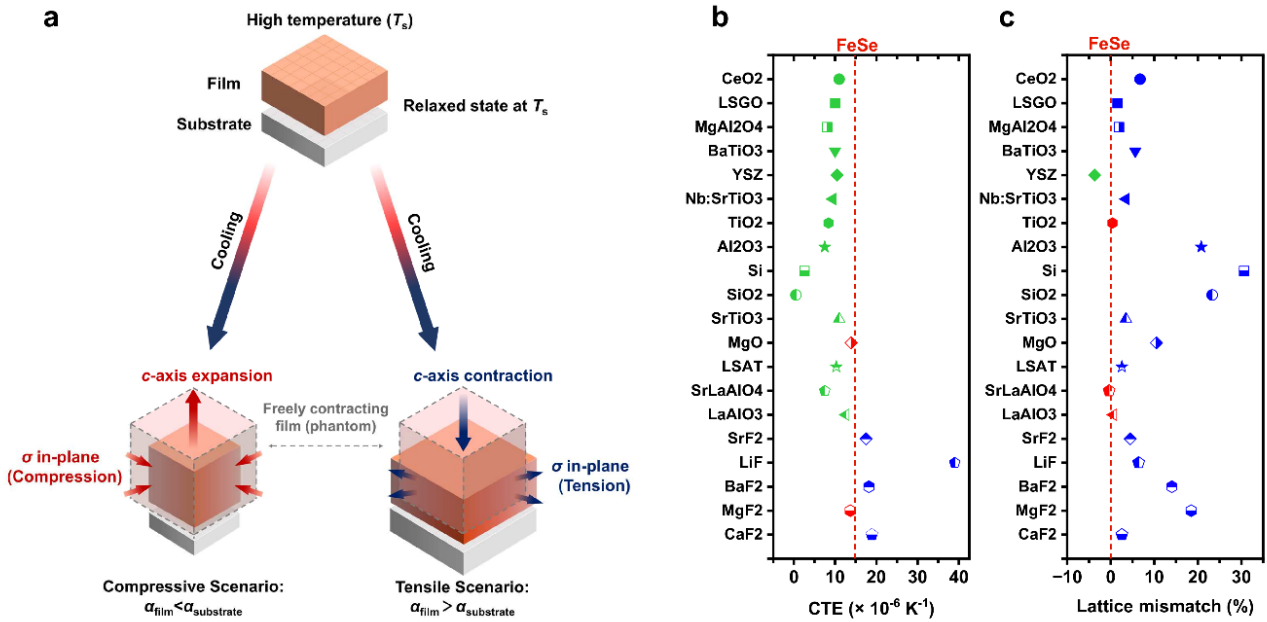

**Fig. S6. Mechanism of in-plane strain and high $T_c$.** (a) Schematic illustration of strain evolution in FeSe films arising from thermal expansion coefficient mismatch with the substrate. At the growth temperature $T_s$, FeSe film is assumed to be fully relaxed on the substrate. Upon cooling, differential contraction occurs: for substrates with $\alpha_{film} < \alpha_{substrate}$ (compressive scenario), the larger contraction of substrate as compared to the film imposing an in-plane compressive stress $\sigma_{\text{in-plane}}$ on the film and resulting in a Poisson-driven $c$-axis expansion. Conversely, for $\alpha_{film} > \alpha_{substrate}$ (tensile scenario), the film is subjected to in-plane tensile stress and a $c$-axis contraction. (b) Comparison of the thermal expansion coefficients of various substrates and FeSe, with the dashed line indicating the CTE of FeSe. Sr/Li/Ba/Ca-F serves as potential substrate to induce compressive strain for FeSe film. (c) Lattice mismatch between FeSe and various substrates, which critically impacts the residual strain and the resulting film properties. Combined bo th CTE and lattice mismatch difference, Sr/Li/Ca-F are considered as the most suitable substrates for realizing high-$T_c$ FeSe films.

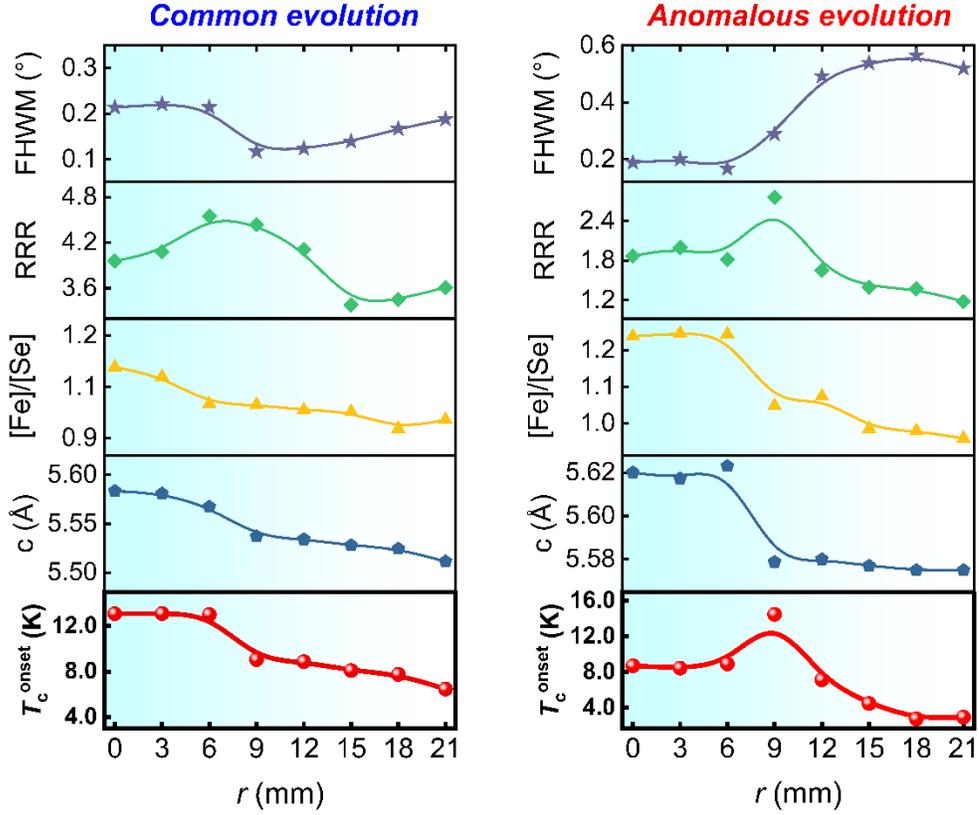

**Fig. S7. Representative radial evolution patterns in combinatorial FeSe films.** Shown are two representative classes of radial evolution observed in off-center-grown FeSe films: a common evolution (left) and an anomalous evolution (right). From top to bottom, the panels plot the x-ray diffraction peak width (FWHM) of (001) peak, residual resistivity ratio (RRR), Fe/Se ratio, $c$-axis lattice parameter, and superconducting onset transition temperature $T_c^{onset}$ as a function of radial position $\boldsymbol{r}$. In the common case, these descriptors evolve relatively smoothly with increasing $r$, and $T_c^{onset}$ decreases monotonically away from the plume center. In the anomalous case, stronger radial variations in crystallinity, stoichiometry, and lattice parameter produce a nonmonotonic $T_c^{onset}(r)$, with the superconducting maximum shifted away from the center. These contrasting behaviors illustrate the competition among strain, stoichiometry, and disorder in determining superconductivity in thick FeSe films.

**Table S1. Comprehensive FeSe film sample library (N = 80) and associated structural, compositional, growth, and transport parameters.** This table summarizes all FeSe films investigated in this work, including the sample index, out-of-plane lattice constant $c$ (Å), composition [Fe]/[Se], superconducting transition temperatures $T_c^{onset}$ and $T_c^{zero}$ (K), residual resistivity ratio (RRR), substrate type, substrate temperature $T_s$ (°C), and XRD peak width (FWHM, °). The $c$-axis lattice constant was extracted from out-of-plane XRD $(00l)$ reflections using Nelson-Riley extrapolation. The [Fe]/[Se] atomic ratio was determined by EDS. $T_c^{onset}$ is defined as the temperature at which $R(T)$ first deviates from the normal-state behavior upon cooling, and $T_c^{zero}$ is the temperature at which the resistance reaches zero. RRR is defined as $R_{300K}/R_n$, where $R_n$ is the normal-state resistance measured immediately above the superconducting transition. FWHM denotes the full width at half maximum of the FeSe (001) peak obtained from the out-of-plane XRD $\theta$-$2\theta$ scan, reflecting the out-of-plane crystalline quality.

| Sample number | $c$ (Å) | [Fe]/[Se] | $T_c^{onset}$ (K) | $T_c^{zero}$ (K) | $RRR$ | Substrate | $T_s$ (°C) | FWHM (°) |
|---|---|---|---|---|---|---|---|---|
| 1 | 5.62 | 1.24 | 8.71 | 7.70 | 1.87 | CaF$_2$ | 300 | 0.19 |
| 2 | 5.62 | 1.25 | 8.45 | 7.27 | 2.00 | CaF$_2$ | 300 | 0.20 |
| 3 | 5.62 | 1.24 | 8.92 | 7.63 | 1.82 | CaF$_2$ | 300 | 0.17 |
| 4 | 5.58 | 1.05 | 14.47 | 12.99 | 2.76 | CaF$_2$ | 300 | 0.29 |
| 5 | 5.58 | 1.07 | 7.14 | 6.09 | 1.65 | CaF$_2$ | 300 | 0.49 |
| 6 | 5.58 | 0.99 | 4.48 | 3.32 | 1.40 | CaF$_2$ | 300 | 0.54 |
| 7 | 5.58 | 0.98 | 2.78 | 1.76 | 1.37 | CaF$_2$ | 300 | 0.56 |
| 8 | 5.58 | 0.96 | 2.97 | 1.76 | 1.18 | CaF$_2$ | 300 | 0.52 |
| 9 | 5.52 | 1.02 | 8.19 | 7.13 | 4.11 | CaF$_2$ | 400 | 0.12 |
| 10 | 5.52 | 1.01 | 6.43 | 5.52 | 3.34 | CaF$_2$ | 400 | 0.11 |
| 11 | 5.51 | 1.01 | 6.17 | 5.05 | 3.03 | CaF$_2$ | 400 | 0.12 |
| 12 | 5.50 | 1.00 | 6.68 | 5.18 | 2.69 | CaF$_2$ | 400 | 0.13 |
| 13 | 5.49 | 0.99 | 3.33 | 0 | 2.64 | CaF$_2$ | 400 | 0.13 |
| 14 | 5.49 | 0.98 | 0 | 0 | 2.28 | CaF$_2$ | 400 | 0.15 |
| 15 | 5.49 | 0.98 | 3.93 | 2.37 | 3.86 | CaF$_2$ | 400 | 0.21 |
| 16 | 5.48 | 0.96 | 0 | 0 | 1.83 | CaF$_2$ | 400 | 0.22 |
| 17 | 5.47 | 0.99 | 0 | 0 | 1.55 | CaF$_2$ | 400 | 0.29 |
| 18 | 5.52 | 1.02 | 9.30 | 8.26 | 4.57 | CaF$_2$ | 500 | 0.13 |
| 19 | 5.52 | 1.00 | 8.97 | 8.04 | 4.37 | CaF$_2$ | 500 | 0.13 |
| 20 | 5.51 | 1.01 | 8.51 | 7.65 | 3.88 | CaF$_2$ | 500 | 0.14 |
| 21 | 5.51 | 0.98 | 6.74 | 5.73 | 3.92 | CaF$_2$ | 500 | 0.14 |
| 22 | 5.50 | 0.99 | 5.43 | 4.08 | 3.68 | CaF$_2$ | 500 | 0.15 |
| 23 | 5.48 | 0.98 | 2.72 | 1.30 | 2.86 | CaF$_2$ | 500 | 0.18 |
| 24 | 5.47 | 0.98 | 0 | 0 | 2.47 | CaF$_2$ | 500 | 0.22 |
| 25 | 5.46 | 0.98 | 0 | 0 | 2.29 | CaF$_2$ | 500 | 0.27 |
| 26 | 5.58 | 1.12 | 13.25 | 12.34 | \ | CaF$_2$ | 400 | 0.22 |
| 27 | 5.58 | 1.12 | 13.16 | 12.28 | \ | CaF$_2$ | 400 | 0.22 |
| 28 | 5.58 | 1.07 | 13.26 | 12.39 | \ | CaF$_2$ | 400 | 0.23 |
| 29 | 5.58 | 1.09 | 13.28 | 12.45 | \ | CaF$_2$ | 400 | 0.24 |
| 30 | 5.55 | 1.07 | 13.33 | 9.55 | \ | CaF$_2$ | 400 | 0.20 |
| 31 | 5.55 | 1.05 | 9.04 | 8.13 | \ | CaF$_2$ | 400 | 0.12 |
| 32 | 5.52 | 1.01 | 8.45 | 7.15 | \ | CaF$_2$ | 400 | 0.14 |
| 33 | 5.50 | 1.00 | 7.75 | 4.13 | \ | CaF$_2$ | 400 | 0.16 |
| 34 | 5.49 | 0.99 | 5.17 | 2.60 | \ | CaF$_2$ | 400 | 0.18 |
| 35 | 5.58 | 1.11 | 13.09 | 12.08 | 3.96 | CaF$_2$ | 500 | 0.21 |
| 36 | 5.58 | 1.08 | 13.10 | 12.07 | 4.08 | CaF$_2$ | 500 | 0.22 |
| 37 | 5.57 | 1.00 | 13.02 | 12.04 | 4.55 | CaF$_2$ | 500 | 0.21 |
| 38 | 5.54 | 1.00 | 9.05 | 8.37 | 4.44 | CaF$_2$ | 500 | 0.12 |
| 39 | 5.53 | 0.98 | 8.87 | 8.08 | 4.11 | CaF$_2$ | 500 | 0.12 |
| 40 | 5.53 | 0.98 | 8.09 | 6.68 | 3.38 | CaF$_2$ | 500 | 0.14 |
| 41 | 5.52 | 0.93 | 7.76 | 6.45 | 3.45 | CaF$_2$ | 500 | 0.17 |
| 42 | 5.51 | 0.95 | 6.47 | 4.66 | 3.61 | CaF$_2$ | 500 | 0.19 |
| 43 | 5.57 | 0.75 | 9.56 | 8.83 | 3.85 | LaAlO$_3$ | 500 | 0.22 |
| 44 | 5.54 | 0.72 | 9.57 | 8.16 | 4.62 | LaAlO$_3$ | 500 | 0.11 |
| 45 | 5.53 | 0.72 | 6.73 | 5.75 | 4.66 | LaAlO$_3$ | 500 | 0.10 |
| 46 | 5.52 | 0.71 | 6.14 | 5.17 | 4.59 | LaAlO$_3$ | 500 | 0.12 |
| 47 | 5.52 | 0.71 | 4.70 | 3.63 | 3.50 | LaAlO$_3$ | 500 | 0.13 |
| 48 | 5.51 | 0.69 | 4.02 | 2.81 | 3.20 | LaAlO$_3$ | 500 | 0.15 |
| 49 | 5.51 | 0.72 | 3.98 | 2.62 | 3.38 | LaAlO$_3$ | 500 | 0.18 |
| 50 | 5.51 | 0.73 | 3.53 | 2.4 | 3.43 | LaAlO$_3$ | 500 | 0.20 |
| 51 | 5.53 | 1.02 | 9.11 | 7.44 | 3.35 | MgO | 500 | 0.15 |
| 52 | 5.53 | 0.98 | 6.06 | 4.60 | 3.27 | MgO | 500 | 0.11 |
| 53 | 5.52 | 0.90 | 4.49 | 3.33 | 3.16 | MgO | 500 | 0.12 |
| 54 | 5.52 | 0.96 | 4.03 | 2.93 | 3.15 | MgO | 500 | 0.12 |
| 55 | 5.51 | 0.97 | 2.90 | 0 | 2.53 | MgO | 500 | 0.12 |
| 56 | 5.51 | 0.96 | 2.20 | 0 | 2.54 | MgO | 500 | 0.13 |
| 57 | 5.51 | 0.91 | 0 | 0 | 2.31 | MgO | 500 | 0.14 |
| 58 | 5.50 | 0.97 | 0 | 0 | 2.39 | MgO | 500 | 0.16 |
| 59 | 5.49 | 1.43 | 3.32 | 0 | 0.85 | Si | 500 | 0.12 |
| 60 | 5.49 | 1.45 | 4.00 | 0 | 0.70 | Si | 500 | 0.12 |
| 61 | 5.48 | 1.50 | 3.35 | 0 | 0.83 | Si | 500 | 0.13 |
| 62 | 5.48 | \ | 2.89 | 0 | 0.48 | Si | 500 | 0.13 |
| 63 | 5.48 | 1.56 | 0 | 0 | 0.51 | Si | 500 | 0.15 |
| 64 | 5.48 | 1.61 | 0 | 0 | 0.39 | Si | 500 | 0.18 |
| 65 | 5.53 | 1.02 | 9.49 | 7.56 | 3.02 | SrTiO$_3$ | 500 | 0.17 |
| 66 | 5.53 | 1.00 | 8.07 | 4.65 | 2.99 | SrTiO$_3$ | 500 | 0.12 |
| 67 | 5.52 | 0.93 | 5.61 | 4.17 | 3.09 | SrTiO$_3$ | 500 | 0.11 |
| 68 | 5.51 | 1.01 | 3.31 | 2.02 | 2.44 | SrTiO$_3$ | 500 | 0.12 |
| 69 | 5.51 | 0.98 | 3.27 | 1.83 | 2.4 | SrTiO$_3$ | 500 | 0.11 |
| 70 | 5.50 | 0.99 | 2.03 | 0 | 2.15 | SrTiO$_3$ | 500 | 0.12 |
| 71 | 5.50 | 0.90 | 2.88 | 1.06 | 2.34 | SrTiO$_3$ | 500 | \ |
| 72 | 5.50 | 0.99 | 0 | 0 | 2.11 | SrTiO$_3$ | 500 | 0.16 |
| 73 | 5.58 | 1.07 | 11.30 | 9 | 2.65 | CaF$_2$ | 500 | 0.34 |
| 74 | 5.59 | 1.00 | 13.00 | 11.80 | 2.81 | CaF$_2$ | 500 | 0.21 |
| 75 | 5.59 | 0.97 | 12.90 | 11.70 | 3.41 | CaF$_2$ | 500 | 0.21 |
| 76 | 5.58 | 1.00 | 12.60 | 11.40 | 2.71 | CaF$_2$ | 500 | 0.32 |
| 77 | 5.57 | 1.01 | 12.48 | 11.52 | 3.95 | CaF$_2$ | 500 | 0.36 |

| 78 | 5.53 | 1.01 | 8.87 | 7.77 | 4.28 | CaF$_2$ | 600 | 0.15 |
| 79 | 5.54 | 0.92 | 8.43 | 6.86 | 1.96 | CaF$_2$ | 450 | 0.34 |
| 80 | 5.61 | 1.01 | 17.10 | 15.20 | 2.80 | CaF$_2$ | 350 | 0.33 |